\documentclass[lettersize,journal]{IEEEtran}
\IEEEoverridecommandlockouts

\usepackage{bbm}
\usepackage{graphicx}

\usepackage{xcolor}
\usepackage{float}   
\usepackage[normalem]{ulem}
\usepackage{dsfont}
\usepackage{textcomp}
\usepackage{stfloats}
\usepackage{url}
\usepackage{verbatim}
\usepackage{algorithmic}
\usepackage{array}
\usepackage{booktabs}
\usepackage{pifont}
\usepackage{comment}
\usepackage{algorithm}
\usepackage{algorithmic}
\usepackage[acronym]{glossaries}
\newacronym{ack}{ACK}{Acknowledgement}
\newacronym{awgn}{AWGN}{Additive white Gaussian Noise}
\newacronym{agv}{AGV}{Automated Guided Vehicle}
\newacronym{ai}{AI}{Artificial Intelligence}
\newacronym{aoi}{AoI}{Age of Information}
\newacronym{bcwu}{BCWu}{BroadCast Wake-up}
\newacronym{bcwuid}{BCWuID}{BroadCast Wake-up ID}
\newacronym{bctiwu}{BCTIWu}{Broadcast TI Wake-up}
\newacronym{csma}{CSMA}{Carrier Sense Multiple Access}
\newacronym{cowu}{\textrm{CoWu}}{Content-based Wake-up}
\newacronym{corwu}{\textrm{CoRWu}}{Content Relevance based Wake-up}
\newacronym{cormo}{CoR-Mode}{Content Relevance-Mode}
\newacronym{cormosg}{CoRMoSg}{CoR-Mode Signal}
\newacronym{cts}{CTS}{Clear to Send}
\newacronym{dl}{DL}{Deep Learning}
\newacronym{dnn}{DNN}{Deep Neural Networks}
\newacronym{dram}{DRAM}{Dynamic Random Access Memory}
\newacronym{es}{ES}{Edge Server}
\newacronym{fp}{FP}{False Positive}
\newacronym{fn}{FN}{False Negative}

\newacronym{id}{ID}{Identity}
\newacronym{idwu}{IDWu}{Identity-based Wake-up}
\newacronym{iot}{IoT}{Internet of Things}

\newacronym{kpis}{KPIs}{Key Performance Indicator}
\newacronym{mcu}{MCU}{Micro Controller Unit}
\newacronym{mac}{MAC}{Medium Access Control}
\newacronym{muac}{MUAC}{Multiply-accumulation}
\newacronym{mcwu}{MCWu}{MultiCast Wake-up}
\newacronym{madd}{mADD}{multiplication and addition}
\newacronym{ml}{\textrm{ML}}{Machine Learning}
\newacronym{mlmo}{ML-Mode}{\gls{ml} Sharing Mode}
\newacronym{mlmosg}{MLMoSg}{\gls{mlmo} Signal}
\newacronym{ook}{OOK}{On-Off Keying}
\newacronym{pmf}{PMF}{Probability Mass Function}
\newacronym{qnns}{QNNs}{Quantized Neural Networks}

\newacronym{proposed2}{\textrm{TinyAirNet}}{Tiny Neural Network transmission over the Air}
\newacronym{ri}{RI}{Receiver-Initiated}
\newacronym{rimosg}{RIMoSg}{RI-Mode Signal}
\newacronym{semdas}{SEMDAS}{SEMantic DAta Soucing}
\newacronym{sram}{SRAM}{Static Random Access Memory}
\newacronym{ti}{TI}{Transmitter-Initiated}
\newacronym{timosg}{TIMoSg}{TI-Mode Signal}

\newacronym{tp}{TP}{True Positive}
\newacronym{tn}{TN}{True Negative}
\newacronym{jscc}{JSCC}{Joint Source Channel Coding}

\newacronym{ucwu}{UCWu}{UniCast Wake-up}
\newacronym{uctiwu}{UCTIWu}{Unicast TI Wake-up}

\newacronym{wsn}{WSNs}{Wireless Sensor Networks}
\usepackage{amsmath,amsfonts, amssymb}
\usepackage{bm}
\usepackage{graphicx}

\usepackage[font=small]{caption}
\usepackage[font=footnotesize]{subcaption}
\usepackage{tikz}
\usetikzlibrary{plotmarks,patterns,decorations.pathreplacing,backgrounds,calc,arrows,arrows.meta,spy,matrix}
\usepackage{tikzscale}

\usepackage{pgfplots}
\usepgfplotslibrary{colorbrewer, groupplots, patchplots}
\pgfplotsset{
    compat=newest,
    legend style={font=\footnotesize, fill opacity=0.7,  draw opacity=1, text opacity=1, draw=white!15!black, legend cell align=left, align=left}, 
    width=0.8\columnwidth, 
    scale only axis,
    height=4cm,
    yminorticks=false,
    xminorticks=false,
    label style={font=\small},
    title style={font=\small},
    tick align=outside,
    tick pos=left,
    tick style={color=black},
    tick label style={font=\footnotesize},
    grid style={line width=.1pt, draw=gray!20},
    major grid style={line width=.1pt,draw=gray!20},
    plot coordinates/math parser=false 
}
\newlength\figureheight
\newlength\figurewidth

\usepackage{multirow}
\usepackage{tablefootnote}
\usepackage{booktabs}
\usepackage{tabularx}

\def\BibTeX{{\rm B\kern-.05em{\sc i\kern-.025em b}\kern-.08em T\kern-.1667em\lower.7ex\hbox{E}\kern-.125emX}}

\begin{document}

\title{TinyAirNet: TinyML Model Transmission for Energy-efficient Image Retrieval from IoT Devices}
\author{Junya Shiraishi,~\IEEEmembership{Member,~IEEE},~Mathias~Thorsager,
~Shashi Raj Pandey,~\IEEEmembership{Member,~IEEE}, \\and~Petar Popovski,~\IEEEmembership{Fellow,~IEEE}\\
\thanks{This work was supported partly by the Villum Investigator Grant ``WATER" from the Velux Foundation, Denmark, and partly by the Horizon Europe SNS ``6G-XCEL" project with Grant 101139194. (Corresponding author is Junya Shiraishi)}
\thanks{J.~Shiraishi, M.~Thorsager, S.~R.~Pandey, and P.~Popovski are with the Department of Electronic Systems, Aalborg University, 9220 Aalborg, Denmark (e-mail: $\{$jush, mdth, srp, petarp$\}$@es.aau.dk)}}




\maketitle

\begin{abstract}
This letter introduces an energy-efficient pull-based data collection framework for Internet of Things (IoT) devices that use Tiny Machine Learning (TinyML) to interpret data queries. A TinyML model is transmitted from the edge server to the IoT devices. The devices employ the model to facilitate the subsequent semantic queries. This reduces the transmission of irrelevant data, but receiving the ML model and its processing at the IoT devices consume additional energy. We consider the specific instance of image retrieval in a single device scenario and investigate the gain brought by the proposed scheme in terms of energy efficiency and retrieval accuracy, while considering the cost of computation and communication, as well as memory constraints. Numerical evaluation shows that, compared to a baseline scheme, the proposed scheme reaches up to $67\%$ energy reduction under the accuracy constraint when many images are stored. Although focused on image retrieval, our analysis is indicative of a broader set of communication scenarios in which the preemptive transmission of an ML model can increase communication efficiency. 

\end{abstract}

\begin{IEEEkeywords}
6G IoT networks, TinyML model, energy efficiency, wireless image retrieval, semantic query
\end{IEEEkeywords}

\section{Introduction}
\begin{figure}[t]
\centering
\includegraphics[width=0.48\textwidth]{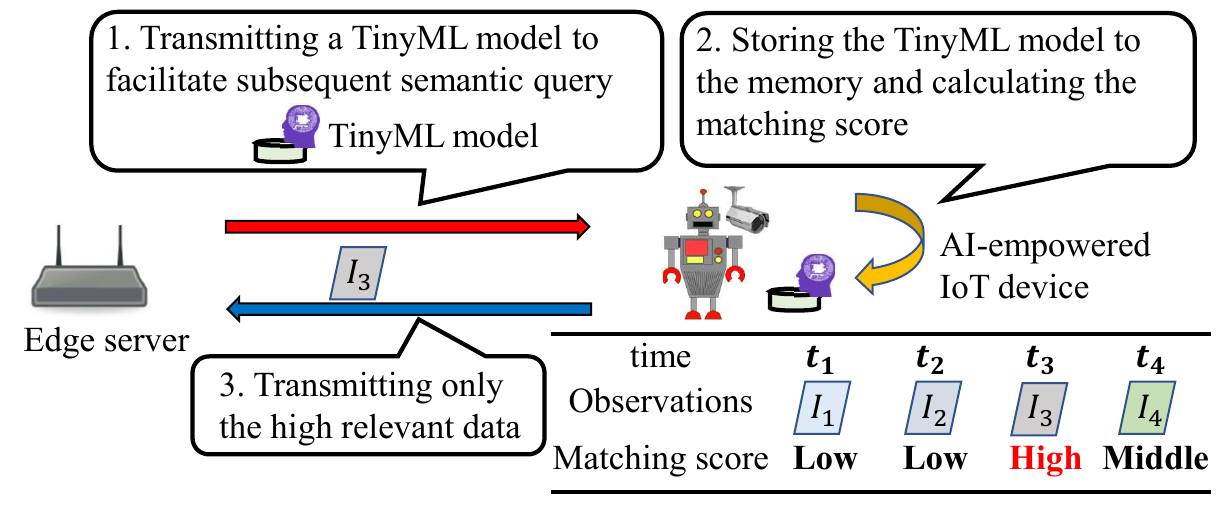}
\caption{An example of \gls{proposed2} for \gls{ai}-empowered \gls{iot} data collection.}
\vspace{-4mm}
\label{Fig:toy_example}
\end{figure}

\IEEEPARstart{E}{nergy} efficiency is one of the essential \gls{kpis} for \gls{iot} networking of sixth generation (6G) communication systems \cite{sheth2020taxonomy,strinati20216g}, in which \gls{ai} technology plays an important role in supporting new emerging applications, including extended reality, connected robotics, and automated systems~\cite{chowdhury20206g}. 
As indiscriminate collection of data may lead to wasteful energy consumption for the \gls{iot} devices, an \gls{ml} model should be introduced in this aspect to filter out irrelevant data to the current query. Then, it becomes crucial to design an energy-efficient communication protocol for \gls{iot} devices serving 6G applications, in which one needs to consider the energy cost caused by the introduction of \gls{ai}/\gls{ml} models as well as that of the primary radio circuit. 
In this context, a framework called \gls{semdas} introduced in~\cite{huang2023semantic} is an attractive approach, in which only the relevant data can be collected by broadcasting the semantic query to the \gls{iot} devices and by calculating a matching score with the help of an \gls{ml} model~\cite{huang2023semantic, kalor2023random}. 
Further, it is important to consider the practical constraints of \gls{iot} devices, such as the memory size, where the typical \gls{mcu} for \gls{iot} devices have extremely limited on-chip \gls{sram} memory ($<$ 512KB) and flash storage ($<$2MB)~\cite{soro2021tinyml, abadade2023comprehensive}. To solve this problem, this letter proposes the transmission of the Tiny\gls{ml} model~\cite{dutta2021tinyml} from the \gls{es} to the \gls{iot} devices considering the query content and timing, which we call \gls{proposed2}. Specific designs for Tiny\gls{ml} include MCUNet~\cite{lin2020mcunet}, EtinyNet~\cite{xu2022etinynet}, etc.
Fig.~\ref{Fig:toy_example} shows an example of \gls{proposed2} for \gls{ai}-empowered data collection. In our framework, first, the \gls{es} transmits a Tiny\gls{ml} model relating to the task of facilitating subsequent semantic query to the \gls{iot} devices, as shown in Fig.~\ref{Fig:toy_example}-1. The \gls{iot} device that receives this Tiny\gls{ml} model stores it in the memory and exploits it to calculate the matching score (Fig.~\ref{Fig:toy_example}-2). After the processing using the Tiny\gls{ml} model, each \gls{iot} device decides whether it transmits data or not to the \gls{es} based on the matching score. In the example of Fig.~\ref{Fig:toy_example}, the \gls{iot} device only transmits data $I_{3}$ whose level of relevance (matching score) is high and suppresses the others (Fig.~\ref{Fig:toy_example}-3), by which the \gls{iot} device can save wasteful energy consumption. From these observations, we can clearly see the advantage of introducing the Tiny\gls{ml} model in terms of energy reduction for data transmission and new challenges in energy-efficient protocol designs, in which we need to consider the additional cost caused by the introduction of the Tiny\gls{ml} model. 
The \gls{proposed2} can be applicable to a variety of \gls{iot} data collection scenarios to reduce wasteful energy consumption, including object detection and vehicles at the edge~\cite{jankowski2020wireless}, \gls{ai} empowered \gls{iot} sensing~\cite{huang2023semantic}, and wild-life animal monitoring/tracking~\cite{elias2017s}. The use case also includes the distributed implementation of the \gls{ml} model as considered in the edge learning~\cite{xu2023edge,letaief2021edge}, in which the \gls{ml} or Tiny\gls{ml} model is deployed over the edge network/device, and the edge device uploads/downloads the model to realize goal-oriented and semantic communication, etc. Furthermore, it could contribute to managing wireless access of \gls{iot} devices belonging to different communication classes, such as the coexistence of pull-based and push-based devices~\cite{Sara_Pull_and_push}. Our work is related to the \gls{semdas}~\cite{huang2023semantic}, but we focus on the energy efficiency of the \gls{iot} data collection and rely on the Tiny\gls{ml} model to send the semantic query. 

The \gls{ml} model transmission over the wireless channel has been considered in the edge networks~\cite{jankowski2024airnet} and in \gls{iot} networks in the context of semantic communication~\cite{xie2020lite}. In \cite{kim2023green}, the authors investigated the energy cost of calculation using \gls{ml} model with \gls{qnns} in a federated learning setup. Moreover, \gls{ml} model transmission has also been considered in 3GPP in the context of \gls{ai}/\gls{ml} model/data distribution and sharing over 5G system~\cite{3GPP_ML} with actual use cases/applications. 
To the best of our knowledge, this is the first work that proposes transmission of Tiny\gls{ml} model to facilitate subsequent semantic query for the wireless image retrieval task~\cite{jankowski2020wireless} in \gls{iot} networks and demonstrates its effectiveness in terms of total energy consumption of \gls{iot} devices and retrieval accuracy. The main contribution of this work is TinyAirNet, an energy-efficient pull-based IoT image retrieval framework, and its associated protocol. The results show significant gains brought by the use of the Tiny\gls{ml} model at the devices in terms of retrieval accuracy and total energy consumption, considering the communication/computation/memory costs.

\section{The \gls{proposed2} Framework}\label{sec: proposed_scheme}

\subsection{Scenario}\label{sec:sys}
 We consider a scenario where an \gls{es} collects data from a single \gls{iot} device, such as a mobile robot, at the specific time instance based on the query requested by a user\footnote{Our proposed framework can be directly applicable to the multi-device setting, which will be kept for future work.}. A mobile robot patrols a given area with a predetermined route and senses the environment with a camera at a predefined interval. The sensing data captured by the robot are stored in a storage $\mathcal{D} = \{\bm{x}_1, \ldots, \bm{x}_{N}\}$. Here, $N$ denotes the number of stored images and $\bm{x}_{j} \in \mathbb{R}^{M_{C}{\times}M_{H}{\times}M_{W}}$ is the $j$-th stored image, where $M_{C}$ is number of channels, $M_{H}$ and $M_{W}$ are the height and width of images in pixel. The communication is conducted in pull-based manners, in which the \gls{es} first transmits the query to retrieve images with a specific object or entity. The mobile robot receiving its query transmits stored images to the \gls{es} through a shared wireless link. We assume that the downlink/uplink channel is error-free.

\subsection{Overview of \gls{proposed2}}
\Gls{proposed2} consists of three phases, as summarized below. 
\paragraph{Tiny\gls{ml} Model Transmission Phase} In this phase, the \gls{es} transmits the Tiny\gls{ml} model related to the task to the \gls{iot} devices. Let $\mathcal{M}$ be the \gls{ml} model that the \gls{es} uses to conduct the task requested by an external user, in which $\mathcal{M}$ is assumed to be the feature extractor for the specific task. Then, in order to share this model with the memory-constrained \gls{iot} devices, the \gls{es} first compresses $\mathcal{M}$ to $\mathcal{\hat{M}}$. For instance, one can apply a neural network transmission strategy as in~\cite{jankowski2024airnet}, in which the \gls{es} compresses the model with a pruning and knowledge distillation considering the communication constraint. The \gls{iot} device receiving the Tiny\gls{ml} model stores the parameter of its model into memory.

\paragraph{Image Retrieval Phase} In order to collect relevant images from the mobile robot, first, the \gls{es} transmits the threshold of similarity measure, $V_{\mathrm{th}}$. The node receiving this information starts processing the observed data using the received Tiny\gls{ml} model, i.e., feature extraction and similarity measure calculation, e.g., calculation of cosine similarity and checks whether the similarity measure of each image exceeds a threshold $V_{\mathrm{th}}$ or not. Note that, as we mentioned, we assume the shared Tiny\gls{ml} is a feature extractor, which outputs a feature vector $\bm{y}$ of an image $\bm{x}$, as ${\mathcal{\hat{M}}}: \bm{x}~\rightarrow~\bm{y}$. Let $g(\cdot, \cdot)$ be the function to extract the similarity measure of an image. Then, the similarity measure of the $j$-th image $\bm{x}_j$ observed by the \gls{iot} device, denoted as $z_{j}~\in~[0, 1]$ is $z_{j}~=~g(\mathcal{\hat{M}}(\bm{x}_j), \mathcal{\hat{M}}(\bm{q}))$, where $\bm{q}$ is a query image. Here, the subset of relevant images extracted by the Tiny\gls{ml} model can be described as follows:
\begin{equation}
\mathcal{\bm{R}}= \{\bm{x}_{j}| z_{j}~\geq V_{\mathrm{th}}, \forall \bm{x}_{j} \in {\mathcal{D}}\}.\label{eq:R}
\end{equation}
 If the \gls{ml} model only has a smaller capability, the observed similarity value $z_{j}$ is highly likely to deviate from the true one, denoted as $\beta_{j} =g\left(\mathcal{{M}}(\bm{x}_j), \mathcal{M}(\bm{q})\right)~\in~[0, 1]$. 
For example, the top-1 accuracy of ImageNet decreases as the model size or quantization value becomes smaller~\cite{xu2022etinynet, han2015deep}. In order to take this into account, we model $z_{j}$ as 
\begin{equation}
z_j = \beta_j + w_{j},\label{eq:observation_model}
\end{equation}
where $w_{j}$ is the observation noise, which follows $w_{j}  \sim \mathcal{N}(0, \sigma_{\mathrm{ML}}^2)$. Here, $\sigma_{\mathrm{ML}}$ is a standard deviation representing the model noise. Considering the trade-off between the \gls{ml} model's size with different quantization levels and its accuracy, we set $\sigma_{\mathrm{ML}}= \frac{1}{b_{q}}$, where ${b_{q}}$ is the number of bits for the quantization of weights. The smaller (larger) value of $\sigma_{\mathrm{ML}}$ represents the high (low) bit quantization, by which the extracted feature of the \gls{ml} model includes less (more) quantization error.
\paragraph{Relevant Image Identification Phase}
After collecting images from the \gls{iot} device, the \gls{es} identifies the relevant images to the query by exploiting the larger \gls{ml} model. Specifically, the \gls{es} first conducts feature extraction for each received image $\hat{\bm{x}}$ using the large \gls{ml} model, whose extracted feature vectors are denoted as $\mathcal{M}(\hat{\bm{x}})$, then calculates the similarity measure by comparing $\mathcal{M}(\hat{\bm{x}})$ with the feature vector of query image $\mathcal{M}(\bm{q})$. The \gls{es} considers the received image as relevant if the similarity measure $\chi_{D} = g\left(\mathcal{M}(\hat{\bm{x}}\right), \mathcal{M}(\bm{q}))$ is higher than the predetermined threshold $\delta$, i.e., $\chi_{D} \geq \delta$. 

Then, we define the retrieval accuracy $\gamma$, as the probability that the estimated relevant image set is the exact ones that are relevant to the query images. Formally, denoting by $\mathcal{T}$ the subset of images whose true similarity measure is higher than $\delta$, and by $\mathcal{S}$ the subset of relevant images that are successfully received by the \gls{es}, then the retrieval accuracy $\gamma$ is:
\begin{equation}
\ \gamma=\mathrm{Pr}(\mathcal{T}=\mathcal{S}).\label{eq:gamma}
\end{equation}
The retrieval accuracy $\gamma$ becomes larger by collecting more images from the mobile robot; however, total energy consumption increases. The goal of this work is to elicit the gain of Tiny\gls{ml} model introduction in terms of energy efficiency and retrieval accuracy.

\section{Energy Model}
\subsubsection{Computation}\label{sec:comp_model}
In this analysis, we assume \gls{qnns} \cite{moons2017minimum}, in which we use only fixed-point representations for both weights and activations for the calculation. The energy cost per inference using the \gls{ml} model can be described as the summation of the energy cost of \gls{dram} ($E_{\mathrm{DRAM}}$) and that of processing at the hardware itself ($E_{\mathrm{HW}}$)~\cite{moons2017minimum}. When the \gls{ml} model can not be stored in the \gls{sram}, we need to rely on off-chip \gls{dram}, in which the larger energy cost is required for data movement \cite{han2015deep}. 
This letter considers the ideal case, where the entire \gls{ml} model can be stored in on-chip memory. With this assumption, $E_{\mathrm{DRAM}}$ can be expressed as the amount of energy consumed to access the input image for \gls{ml} model, as expressed below:
\begin{equation} E_{\mathrm{DRAM}}=E_{D}\times(M_{C} \times M_{H} \times {M_{W}} \times  \frac{b_{\mathrm{in}}}{b_{q}}),\end{equation}
where $E_{D}$ is the energy consumed per int $b_{q}$ \gls{dram} access and $b_{\mathrm{in}}$ is the number of bits for a single pixel~\cite{moons2017minimum}.

The energy consumption for the read/write from/to the small local \gls{sram} or Register file $E_{L}$ is modeled to be equal to the energy of a single \gls{muac} operation $E_{\mathrm{MUAC}}$, while accessing the main \gls{sram} costs $E_{M} = 2E_{\mathrm{MUAC}}$ \cite{moons2017minimum}. 
Then, $E_{\mathrm{HW}}$ can be modeled as the summation of the compute energy ($E_{C}$), cost of the weight ($E_{W}$), and the activation access ($E_{A}$), as follows \cite{moons2017minimum}: 
\begin{equation}
\ E_{\mathrm{HW}} = E_{C} + E_{W} + E_{A},\label{eq:HW}
\end{equation}
with $E_{C}=E_{\mathrm{MUAC}}\times(N_{c}+3\times A_{s})$, $E_{W}=E_{M}\times N_{s}+ E_{L} \times N_{c}/ \sqrt{p}$, and $E_{A}=2\times E_{M}\times A_{s}+E_{L}\times N_{c}/\sqrt{p}$, 
where $N_{c}$ is the network complexity in the number of \gls{muac} operations, $N_{s}$ is the model size in the number of weights and biases, $A_{s}$ is the total number of activations throughout the whole network, and $p = 64 \times {b_{\max}}/{b_{q}}$, where $b_{\max}$ is the full-precision bit.

\subsubsection{Communication}\label{sec:comm_cost}
The \gls{iot} device consumes energy either in the transmit or the receive state. The power consumption at the receive/transmit state is denoted as $\xi_{R}$/$\xi_{T}$ [W]. To simplify the analysis, we ignore the power consumed during idle periods. The energy consumed for receiving data can be calculated by considering the time required for receiving the Tiny\gls{ml} model $t_{\mathrm{ML}}$ and query feature vector $t_{\mathrm{q}}^{\mathrm{F}} $. A model consists of model topology and model weight factors~\cite{3GPP_ML}. Then, the time required for Tiny\gls{ml} model reception can be 
\begin{equation}
t_{\mathrm{ML}} = \frac{N_{s}b_{q} + b_{\mathrm{ml}} + b_{\mathrm{h}}^{\gls{ml}}}{R_{\mathrm{DL}}},
\end{equation}
where $R_{\mathrm{DL}}$ is the transmission rate for the downlink, $b_{\mathrm{ml}}$ is the number of bits required for the information of model/topology, and $b_{\mathrm{h}}^{\gls{ml}}$ is the number of bits for a header in Tiny\gls{ml} transmission. Here, we set $b_{\mathrm{ml}} = 0$, as the size of the model topology is much smaller than the actual model size~\cite{3GPP_ML}. Similarly, the time for reception of the query feature vector can be described as $t_{\mathrm{q}}^{\mathrm{F}} = (l_{\mathrm{F}} + b_{\mathrm{h}}^{\mathrm{F}})/{R_{\mathrm{DL}}}$, where $b_{\mathrm{h}}^{\mathrm{F}}$ is the number of bits for header feature vector transmission and $l_{\mathrm{F}}$ is the size of feature vector. Here, for simplicity, we set $b_{\mathrm{h}}^{\gls{ml}} = 0$ and $b_{\mathrm{h}}^{\mathrm{F}} = 0$. On the other hand, the energy consumed for transmitting data depends on the size of $|\mathcal{\bm{R}}| = \psi$ in Eq.~\eqref{eq:R}.
Then, the total energy consumed for the image retrieval task at the specific time instance from \gls{iot} device can be:
\begin{equation}
\ E_{\mathrm{comm}} = \xi_{R}(t_{\mathrm{ML}} + t_{\mathrm{q}}^{\mathrm{F}}) + \xi_{T} t_{\mathrm{data}}\psi,\label{eq:comm}
\end{equation}
where $t_{\mathrm{data}}$ is the time required for data transmission of a single image and $t_{\mathrm{data}} = ({b_{\mathrm{in}} M_{C} M_{H} {M_{W}}} +b_{\mathrm{h}})/{R_{\mathrm{UL}}}$,
where $R_{\mathrm{UL}}$ is the data rate for uplink transmission.

\section{Analysis}\label{sec:analysis}
We derive the total energy consumption and retrieval accuracy when we apply the \gls{proposed2} for the wireless image retrieval tasks. Here, total energy consumption is defined as the total amount of energy consumed by a single device, while the retrieval accuracy is defined by Eq.~\eqref{eq:gamma}. In this analysis, we ignore the energy consumed for the reception of $V_{\mathrm{th}}$, as it is much smaller, compared with the cost of Tiny\gls{ml} model reception and image transmission.

Based on the energy model in Sec.~\ref{sec:comp_model}, the total energy consumed for computation can be expressed as:
\begin{equation}
E_{\mathrm{comp}}^{\gls{proposed2}} (N) = N\left(E_{\mathrm{DRAM}}+E_{\mathrm{HW}} + n_{\mathrm{F}}E_{\mathrm{MUAC}}\right),\label{eq:energy_ML}
\end{equation}
where $n_{\mathrm{F}}$ is the number of \gls{muac} operations to calculate one similarity measure, which is $n_{\mathrm{F}} = l_{\mathrm{F}}(l_{\mathrm{F}}-1)+2l_{\mathrm{F}}^2 + 2$.
Next, we derive the energy cost for communication based on the model described in Sec.~\ref{sec:comm_cost}. Let $P_{\mathrm{th}} (V_{\mathrm{th}})$ be the probability the similarity measure of an observed image $\bm{x}$ is equal to or higher than the threshold of $V_{\mathrm{th}}$. According to the Eq.~\eqref{eq:observation_model} the conditional probability of $z$ given $\beta$ can be expressed as $p(z|\beta) = \frac{1}{\sqrt{2\pi}\sigma^2}\int_{-\infty}^{\infty}\exp{\left(-\frac{(z-\beta)^2}{2\sigma_{\mathrm{ML}}^2}\right)}dz$. Then, $P_{\mathrm{th}} (V_{\mathrm{th}})$ can be expressed as follows:
 \begin{equation}
 \ P_{\mathrm{th}} (V_{\mathrm{th}}) = \int_{0}^{1}Q(\frac{V_{\mathrm{th}}-\beta}{\sigma_{\mathrm{ML}}})g_{T}(\beta)d{\beta},
 \end{equation}
where $Q(x)$ denotes the Q-function defined as $Q(x) = \frac{1}{\sqrt{2\pi}}\int_{x}^{\infty}\exp(-\frac{u^2}{2})du$ and $g_{T}(\beta)$ is the distribution of true similarity value. Then, the probability of $w$ out of $N$ images satisfying the threshold of $V_{\mathrm{th}}$, follows binomial distribution as:
\begin{equation}
    P_{D} (w)= \binom{N}{w}P_{\mathrm{th}} (V_{\mathrm{th}})^w(1-P_{\mathrm{th}} (V_{\mathrm{th}}))^{N-w}.
\end{equation}
Since the value of $w$ is a random variable, the expected total energy consumption for communication, including the energy cost for image transmission and for Tiny\gls{ml} model and feature vector reception, can be expressed as follows:
\begin{equation}
\ E_{\mathrm{comm}}^{\gls{proposed2}}(N) = \sum_{w=0}^{N}P_{D} (w) w{\xi_T}t_{\mathrm{data} } + (t_{\mathrm{ML}} + t_{\mathrm{q}}^{\mathrm{F}})\xi_{R}.\label{eq:energy_data}
\end{equation}
With Eqs.~\eqref{eq:energy_ML} and \eqref{eq:energy_data}, the total energy consumption of \gls{proposed2} can be expressed as follows:
\begin{equation}
\ E_{\mathrm{total}}^{\gls{proposed2}} (N) =E_{\mathrm{comp}}^{\gls{proposed2}} (N) +E_{\mathrm{Data}}^{\gls{proposed2}}(N).\end{equation}

Now, we derive the retrieval accuracy of \gls{proposed2} $\gamma_{\gls{proposed2}}$ defined by Eq.~\eqref{eq:gamma}. Let us denote the probability that the true similarity measure of an image is equal to or higher than $\delta$ as $P_{\delta}$, which can be expressed as $P_{\delta}=\int_{\delta}^{1}g_{T}(\beta)d{\beta}$. We are then interested in the probability that an actual relevant image is successfully delivered to the \gls{es}, denoted as $P_{A}$, which is $\mathrm{Prob} (z \geq V_{\mathrm{th}}| \beta \geq \delta)$. This can be described as:
\begin{equation}\begin{split}
P_A =\frac{1}{P_{\delta}}\int_{\delta}^{1}{Q\left(\frac{V_{\mathrm{th}}-\beta}{\sigma_{\mathrm{ML}}}\right)}g_{T}(\beta)d{\beta}. \label{eq:pr_A}
\end{split}\end{equation}
Let $\zeta$ be a random variable representing the number of actual relevant images observed by the \gls{iot} device. The probability that $\zeta$ out of $N$ images is actually relevant can be $P_{\zeta}(\zeta) = \binom{N}{\zeta}P_{\delta}^\zeta(1-P_{\delta})^{N-\zeta}$. Then, according to the definition in Eq.~\eqref{eq:gamma}, the probability that the retrieval accuracy is one given $\zeta$ can be described as $P_A^{\zeta}$. 
Finally, the expected retrieval accuracy for \gls{proposed2} can be described as
\begin{equation}
\ \gamma_{\gls{proposed2}}= \sum_{\zeta=0}^{N}P_A^{\zeta}P_{\zeta}(\zeta).
\end{equation}
\section{Numerical Evaluation}
\subsection{Simulation Setting}
 We conduct the simulation based on the description in Sec.~\ref{sec: proposed_scheme}. 
 We set $R_{\mathrm{UL}} =R_{\mathrm{DL}} = 10^5$ [bps], and $b_{\max} = 16$~\cite{moons2017minimum}. Assuming NB-IoT, $\xi_{T}$ and $\xi_{R}$ are set to be 170~mW and 160~mW, respectively~\cite{BG96,khan2021empirical}.
 The values of $E_{\mathrm{MUAC}}$, $E_{L}$, $E_{M}$, and $E_{D}$ are set to $E_{\mathrm{MUAC}}= 3.7 \text{pJ} \times ({b_{q}}/{b_{\max}})^{1.25}$, $E_{L}= 3.7 \text{pJ} \times ({b_{q}}/{b_{\max}})$, $E_{M} = 2 \times 3.7 \text{pJ} \times ({b_{q}}/{b_{\max}})$, and $E_{D}= 128 \times 3.7 \text{pJ} \times ({b_{q}}/{b_{\max}})$ \cite{moons2017minimum}. For simplicity of analysis, we generate $g_{T}(\beta)$ based on uniform distribution with the range of [0, 1]. Here, we select the parameter of Tiny\gls{ml} model from EtinyNet1.0~\cite{xu2022etinynet} as a showcase of performance evaluations, where $N_{s}$ and $N_{c}$ are 0.976 M and 117 M, and we set $A_{s}$ to 4.309 M based on our calculation for the base EtinyNet1.0 architecture. The image size is set to ($M_{C}$, $M_{H}$, $M_{W}$, $b_{\mathrm{in}}$) = (3, 256, 256, 8), considering the STM32F746 MCUs~\cite{xu2022etinynet} and the feature vector is set to $l_{\mathrm{F}} =1000$~based on the size of a fully connected layer of EtinyNet~\cite{xu2022etinynet}. We conducted a simulation $10^4$ times. 
 
 As a baseline scheme, we consider the simple offloading scheme, in which the \gls{iot} device transmits all observations without computations. In this scheme, as the \gls{es} can collect all images without collisions, retrieval accuracy is always 1, while the total energy consumption of the baseline is
\begin{equation}
\ E_{\mathrm{Baseline}} = Nt_{\mathrm{data}}\xi_{T}.
\end{equation}
\subsection{Numerical Results}
\begin{figure}[t!]
\centering
%
%
\definecolor{mycolor1}{rgb}{0.85000,0.32500,0.09800}%
\begin{tikzpicture}

\begin{axis}[%
width=6cm,
height=3.5cm,
scale only axis,
xmin=0.6,
xmax=1,
xlabel style={font=\color{white!15!black}},
xlabel={$\text{V}_{\text{th}}$},
ymin=0,
ymax=30,
every y tick label/.append style={color=blue},
ylabel style={font=\color{blue}},
ylabel={Total energy consumption [J]},
axis y line*=left, 
axis x line*=bottom,
axis lines*=box,
]
\addplot [color=blue, line width=1pt]
  table[row sep=crcr]{%
0.98046875	14.1088995461182\\
0.9609375	14.4183584001712\\
0.94140625	14.7587628010474\\
0.921875	15.1282876782431\\
0.90234375	15.5245573276821\\
0.8828125	15.9447998145743\\
0.86328125	16.386014312307\\
0.84375	16.845134901085\\
0.82421875	17.3191766148639\\
0.8046875	17.8053534016977\\
0.78515625	18.3011623387235\\
0.765625	18.8044330841702\\
0.74609375	19.3133454636133\\
0.7265625	19.8264208355914\\
0.70703125	20.3424943078321\\
0.6875	20.8606750789874\\
0.66796875	21.3803014382195\\
0.6484375	21.9008956246982\\
0.62890625	22.4221221848303\\
0.609375	22.9437519536899\\
0.58984375	23.4656325176547\\
0.5703125	23.9876650763777\\
0.55078125	24.509787019817\\
0.53125	25.0319592202563\\
0.51171875	25.5541569322482\\
0.4921875	26.0763632120552\\
0.47265625	26.5985638390726\\
0.453125	27.1207427894616\\
0.43359375	27.6428773423448\\
0.4140625	28.1649318785495\\
};
\label{plot_one}

\addplot [color=blue, line width=1pt, only marks, mark size=1, mark=o, mark options={solid, blue}]
  table[row sep=crcr]{%
0.98046875	14.094842062536\\
0.9609375	14.457418671816\\
0.94140625	14.7863045342161\\
0.921875	15.1130513015761\\
0.90234375	15.5090512708562\\
0.8828125	15.9095968170961\\
0.86328125	16.387951945416\\
0.84375	16.8267338154959\\
0.82421875	17.3240734122958\\
0.8046875	17.7914656785358\\
0.78515625	18.2593927185358\\
0.765625	18.7987120554957\\
0.74609375	19.2826823082957\\
0.7265625	19.7869739639757\\
0.70703125	20.3276302353357\\
0.6875	20.8426173662157\\
0.66796875	21.3779258999758\\
0.6484375	21.8995977028558\\
0.62890625	22.4972073796559\\
0.609375	22.885453129416\\
0.58984375	23.488410543816\\
0.5703125	24.0210452087761\\
0.55078125	24.5443213329361\\
0.53125	25.0654583620562\\
0.51171875	25.5587871556563\\
0.4921875	26.1138823185362\\
0.47265625	26.5590814737363\\
0.453125	27.1350328132563\\
0.43359375	27.6361158263763\\
0.4140625	28.2318537950165\\
};
\label{plot_two}
\end{axis}

\begin{axis}[%
width=6cm,
height=3.5cm,
 axis lines*=box,
scale only axis,
xmin=0.6,
xmax=1,
xlabel style={font=\color{white!15!black}},
xlabel={$\text{V}_{\text{th}}$},
ymin=0,
ymax=1,
every y tick label/.append style={color=red},
ylabel style={font=\color{red}},
ylabel={Retrieval Accuracy},
axis y line*=right, 
axis x line*=bottom,
legend style={at={(0,0)}, anchor=south west}
],

\addplot [color=red, dashdotted, line width=1pt]
  table[row sep=crcr]{%
0.98046875	0.542161713082159\\
0.9609375	0.577676274592406\\
0.94140625	0.615774618947662\\
0.921875	0.655700066889738\\
0.90234375	0.696520843212077\\
0.8828125	0.737192539827116\\
0.86328125	0.776643134047006\\
0.84375	0.813868292296722\\
0.82421875	0.848021012749961\\
0.8046875	0.878480138700289\\
0.78515625	0.904886950221832\\
0.765625	0.927146361180512\\
0.74609375	0.945396774418382\\
0.7265625	0.95995814282026\\
0.70703125	0.971270060771144\\
0.6875	0.979830863519337\\
0.66796875	0.986145761585533\\
0.6484375	0.990688312232093\\
0.62890625	0.993876137307708\\
0.609375	0.996059364782992\\
0.58984375	0.997518954912078\\
0.5703125	0.998471705988996\\
0.55078125	0.999079014106122\\
0.53125	0.999457076939504\\
0.51171875	0.999686941619129\\
0.4921875	0.999823447009253\\
0.453125	0.999947482344267\\
0.4140625	0.999985720202622\\
};
\label{plot_three}
\addlegendentry{\scriptsize$\text{Retrieval Accuracy (Theory)}$}

\addplot [color=red, line width=1pt, only marks, mark size=1, mark=triangle, mark options={solid, red}]
  table[row sep=crcr]{%
0.98046875	0.5415\\
0.9609375	0.5869\\
0.94140625	0.6136\\
0.921875	0.6615\\
0.90234375	0.7026\\
0.8828125	0.7469\\
0.86328125	0.773\\
0.84375	0.8137\\
0.82421875	0.8498\\
0.8046875	0.8786\\
0.78515625	0.9025\\
0.765625	0.9293\\
0.74609375	0.948\\
0.7265625	0.9591\\
0.70703125	0.9716\\
0.6875	0.9802\\
0.66796875	0.985\\
0.6484375	0.9908\\
0.62890625	0.9949\\
0.609375	0.9953\\
0.58984375	0.9974\\
0.5703125	0.9988\\
0.55078125	0.9992\\
0.53125	0.9995\\
0.51171875	0.9998\\
0.4921875	1\\
0.47265625	0.9999\\
0.453125	0.9998\\
0.43359375	1\\
0.4140625	1\\
};
\label{plot_Four}
\addlegendentry{\scriptsize$\text{Retrieval Accuracy (Simulation)}$}

\addlegendimage{/pgfplots/refstyle=plot_one}\addlegendentry{\scriptsize$\text{Total energy consumption (Theory)}$}
\addlegendimage{/pgfplots/refstyle=plot_two}\addlegendentry{\scriptsize$\text{Total energy consumption (Simulation)}$}
\end{axis}

\end{tikzpicture}%
\caption{Total energy consumption and retrieval accuracy of \gls{proposed2} against the threshold of $V_{\mathrm{th}}$.}
\label{Fig:vth_Trade_off}
\vspace{-4mm}
\end{figure}
Fig.~\ref{Fig:vth_Trade_off} shows the total energy consumption and retrieval accuracy of \gls{proposed2} against threshold of $V_{\mathrm{th}}$, where we set $b_{q} = 8$, $\delta = 0.9$, and $N = 10$. From this figure, first we can see that the results for \gls{proposed2} obtained by theoretical analysis coincide with that of simulation results for both total energy consumption and retrieval accuracy, which validates our analysis. Next, from Fig.~\ref{Fig:vth_Trade_off}, we can see the basic trade-off between the retrieval accuracy and the total energy consumption through the value of $V_{\mathrm{th}}$: the smaller (higher) $V_{\mathrm{th}}$ realizes higher (lower) retrieval accuracy because of the increasing (decreasing) number of available images at the \gls{es}, while it increases (decreases) the energy consumption for image transmission. 

Because of this trade-off, we compare the minimum energy consumption of \gls{proposed2} that achieves the target retrieval accuracy $\gamma_{\mathrm{th}}$ with the baseline. To this end, first, we obtain the optimal parameter of $V_{\mathrm{th}}^{\mathrm{opt}}$ that minimizes the total energy consumption of \gls{proposed2} under the constraint of $\gamma_{\gls{proposed2}} \geq \gamma_{\mathrm{th}}$, as expressed below:
\begin{equation}\begin{split}
\ V_{\mathrm{th}}^{\mathrm{opt}}(N)= &\min_{V_{\mathrm{th}}~\in~[0, 1]}E_{\mathrm{total}}^{\gls{proposed2}}(N)\\&\text{s.t}.~\gamma_{\gls{proposed2}} \geq \gamma_{\mathrm{th}},
\end{split}\end{equation}
where we vary the value of $V_{\mathrm{th}}$ with a step of $\frac{1}{2^8}$. 
Through these evaluations, we set $\gamma_{\mathrm{th}}= 0.98$, $b_{q} =$ 8, 16, and $\delta =$ 0.90, 0.98. Fig.~\ref{Fig:Optimal_vth} shows the optimal threshold of $(V_{\mathrm{th}}^{\mathrm{opt}})$ against $N$. First, we can see that the value of $V_{\mathrm{th}}^{\mathrm{opt}}$ is always equal to or lower than $\delta$, and $V_{\mathrm{th}}^{\mathrm{opt}}$ becomes lower when the value of $b_{q}$ is smaller and/or the value of $N$ is larger. This is because as the value of $b_{q}$ becomes smaller, the observed similarity measure is more likely to deviate from the true one, by which the \gls{es} needs to select the lower value of $V_{\mathrm{th}}$ in order to ensure the retrieval accuracy to exceed the threshold $\gamma_{\mathrm{th}}$. Likewise, as $N$ increases, the probability of these deviations increases, which requires the \gls{es} to set the lower threshold $V_{\mathrm{th}}$.

\begin{figure}[t!]
\centering
\begin{tikzpicture}

\definecolor{crimson2143940}{RGB}{214,39,40}
\definecolor{darkgray176}{RGB}{176,176,176}
\definecolor{darkorange25512714}{RGB}{255,127,14}
\definecolor{forestgreen4416044}{RGB}{44,160,44}
\definecolor{lightgray204}{RGB}{204,204,204}
\definecolor{mediumpurple148103189}{RGB}{148,103,189}

\def\hsep{1.5cm}
\def\vsep{1cm}
\def\vside{2.5cm}
\def\hside{0.3\columnwidth}

\definecolor{mycolor1}{rgb}{0.49000,0.18000,0.56000}%
\begin{groupplot}[
group style={group name=syssys, group size=2 by 1,  horizontal sep=\hsep, vertical sep=\vsep,  xlabels at=edge bottom}, 
anchor=south east,
height=\vside,
width=\hside,
scale only axis,
legend style={  
  at={(-0.37, 1.05)}, 
  legend columns=2,
  draw=none,
  fill opacity=0,
  anchor=south,  
  /tikz/every even column/.append style={column sep=0.3cm}
},
xlabel={$N$},
xmin=0, xmax=50,
xtick={0, 10, 20, 30, 40, 50},
xticklabels={
 \(\displaystyle {0}\),
  \(\displaystyle {10}\),
  \(\displaystyle {20}\),
  \(\displaystyle {30}\), 
  \(\displaystyle {40}\), 
  \(\displaystyle {50}\), 
},]
ymajorgrids,
ylabel shift=-5pt,
xlabel shift=-3pt,
every x tick scale label/.style={at={(rel axis cs:1,0)},anchor=south west,inner sep=0pt, font=\tiny},
]

\nextgroupplot[
ylabel={$V_{\mathrm{th}}$},
ymin=0.6, ymax=1,
ytick={0.5,0.75,1},
yticklabels={
  \(\displaystyle {0.5}\),
  \(\displaystyle {0.75}\),
  \(\displaystyle {1}\),  
},]
\addplot [color=red, line width=1, mark size=1, mark=o, mark options={solid, red}]
  table[row sep=crcr]{%
2	0.78515625\\
4	0.73828125\\
6	0.71484375\\
8	0.6953125\\
10	0.68359375\\
12	0.67578125\\
14	0.66796875\\
16	0.66015625\\
18	0.65625\\
20	0.6484375\\
22	0.64453125\\
24	0.640625\\
26	0.63671875\\
28	0.6328125\\
30	0.6328125\\
32	0.62890625\\
34	0.625\\
36	0.62109375\\
38	0.62109375\\
40	0.6171875\\
42	0.6171875\\
44	0.61328125\\
46	0.61328125\\
48	0.609375\\
50	0.609375\\
};

\addplot [color=blue, line width=1, mark size=1, mark=diamond, mark options={solid, blue}]
  table[row sep=crcr]{%
2	0.98828125\\
4	0.90625\\
6	0.8671875\\
8	0.84375\\
10	0.828125\\
12	0.81640625\\
14	0.8046875\\
16	0.796875\\
18	0.7890625\\
20	0.78125\\
22	0.77734375\\
24	0.7734375\\
26	0.765625\\
28	0.76171875\\
30	0.7578125\\
32	0.75390625\\
34	0.75390625\\
36	0.75\\
38	0.74609375\\
40	0.7421875\\
42	0.7421875\\
44	0.73828125\\
46	0.734375\\
48	0.734375\\
50	0.73046875\\
};

\addplot [color=black!50!green, dashdotted, line width=1, mark size=1, mark=o, mark options={solid, black!50!green}]
  table[row sep=crcr]{%
2	0.859375\\
4	0.8359375\\
6	0.8203125\\
8	0.8125\\
10	0.80859375\\
12	0.80078125\\
14	0.796875\\
16	0.79296875\\
18	0.79296875\\
20	0.7890625\\
22	0.78515625\\
24	0.78515625\\
26	0.78125\\
28	0.78125\\
30	0.77734375\\
32	0.77734375\\
34	0.77734375\\
36	0.7734375\\
38	0.7734375\\
40	0.7734375\\
42	0.76953125\\
44	0.76953125\\
46	0.76953125\\
48	0.76953125\\
50	0.765625\\
};

\addplot [color=mycolor1, dashdotted, line width=1, mark size=1, mark=diamond, mark options={solid, mycolor1}]
  table[row sep=crcr]{%
2	0.98828125\\
4	0.9453125\\
6	0.92578125\\
8	0.91796875\\
10	0.90625\\
12	0.90234375\\
14	0.89453125\\
16	0.890625\\
18	0.88671875\\
20	0.88671875\\
22	0.8828125\\
24	0.87890625\\
26	0.87890625\\
28	0.875\\
30	0.875\\
32	0.87109375\\
34	0.87109375\\
36	0.8671875\\
38	0.8671875\\
40	0.8671875\\
42	0.86328125\\
44	0.86328125\\
46	0.86328125\\
48	0.859375\\
50	0.859375\\
};

\nextgroupplot[
ylabel={$\eta$},
ymin=0.0, ymax=2,
ytick={0, 0.5, 1, 1.5, 2},
yticklabels={
 \(\displaystyle {0}\),
  \(\displaystyle {0.5}\),
  \(\displaystyle {1.0}\),
  \(\displaystyle {1.5}\),  
\(\displaystyle {2}\),
},
]

\addplot [color=red, line width=1, mark size=1, mark=o, mark options={solid, red}]
  table[row sep=crcr]{%
2	2.55355694111557\\
4	1.43087603730789\\
6	1.06457608483017\\
8	0.889228253843097\\
10	0.784051272935732\\
12	0.713943466980858\\
14	0.666093746514074\\
16	0.632156840682364\\
18	0.603601020063153\\
20	0.585432449965318\\
22	0.568089643412562\\
24	0.554287675956083\\
26	0.543209555011539\\
28	0.534271699118738\\
30	0.523146033031991\\
32	0.517311177901859\\
34	0.512622206028969\\
36	0.508888182953606\\
38	0.502056633602095\\
40	0.499809969608967\\
42	0.494247136565593\\
44	0.493092193724765\\
46	0.488474822423941\\
48	0.488144816742315\\
50	0.484250833611953\\
};
\addlegendentry{\scriptsize$\text{TinyAirNet (}\delta\text{ = 0.90, b}_\text{q}\text{ = 8)}$}

\addplot [color=blue, line width=1, mark size=1pt, mark=diamond, mark options={solid, blue}]
  table[row sep=crcr]{%
2	2.39247576814424\\
4	1.27848054854588\\
6	0.920981655060575\\
8	0.746810572795681\\
10	0.644134519999622\\
12	0.577064152033046\\
14	0.532394290634945\\
16	0.498051753113909\\
18	0.473032652406013\\
20	0.454550456810531\\
22	0.437065488583613\\
24	0.423130473316815\\
26	0.415711180208569\\
28	0.406665504634133\\
30	0.39933938672732\\
32	0.393411432413925\\
34	0.38482176374401\\
36	0.38100052018968\\
38	0.377989580626143\\
40	0.375667985932839\\
42	0.370105152889465\\
44	0.368880638118383\\
46	0.368101314410513\\
48	0.363868724051424\\
50	0.363817882863201\\
};
\addlegendentry{\scriptsize$\text{TinyAirNet (}\delta\text{ = 0.98, b}_\text{q}\text{ = 8)}$}

\addplot [color=black!50!green, dashdotted, line width=1, mark size=1, mark=o, mark options={solid, black!50!green}]
  table[row sep=crcr]{%
2	4.81365601177449\\
4	2.50067303679641\\
6	1.73750369607179\\
8	1.35592981127813\\
10	1.12620726856908\\
12	0.978263422808228\\
14	0.870917616301076\\
16	0.791384691022503\\
18	0.726489127626788\\
20	0.678477309402742\\
22	0.63990517438439\\
24	0.604507594350364\\
26	0.578461022451812\\
28	0.552788052317244\\
30	0.53444358529732\\
32	0.514974916278606\\
34	0.497796678909152\\
36	0.486432746523995\\
38	0.472770522651213\\
40	0.460474521165709\\
42	0.453255316413575\\
44	0.443141722118139\\
46	0.433907570804915\\
48	0.425442932101126\\
50	0.421561322519041\\
};
\addlegendentry{\scriptsize$\text{TinyAirNet (}\delta\text{ = 0.90, b}_\text{q}\text{ = 16)}$}

\addplot [color=mycolor1, dashdotted, line width=1, mark size=1, mark=diamond, mark options={solid, mycolor1}]
   table[row sep=crcr]{%
2	4.70399662905074\\
4	2.3977833436707\\
6	1.63559531127762\\
8	1.25320788112095\\
10	1.0303635730189\\
12	0.878274827924153\\
14	0.774430025183116\\
16	0.694731718216561\\
18	0.633596078726781\\
20	0.58167962801021\\
22	0.542981314513569\\
24	0.511379324296044\\
26	0.481427525805715\\
28	0.459565066173395\\
30	0.437315158723436\\
32	0.421670194464934\\
34	0.40449195709548\\
36	0.393057739659149\\
38	0.379395515786367\\
40	0.367099514300863\\
42	0.359820084910481\\
44	0.349706490615045\\
46	0.340472339301821\\
48	0.335862137470212\\
50	0.328074669862726\\
};
\addlegendentry{\scriptsize$\text{TinyAirNet (}\delta\text{ = 0.98, b}_\text{q}\text{ = 16)}$}
\end{groupplot}

\node[text width=\hside, align=center, anchor=north] at ([yshift=-0.8cm]syssys c1r1.south) {\protect\subcaption{$V_{\mathrm{th}}$ against $N$.\protect\label{Fig:Optimal_vth}}};
\node[text width=\hside, align=center, anchor=north] at ([yshift=-0.8cm]syssys c2r1.south) {\protect\subcaption{$\eta$ against $N$.\protect\label{Fig:results_EE}}};

\end{tikzpicture}
\caption{$V_{\mathrm{th}}$ and $\eta$ against $N$.}
\label{Fig:Group}
\vspace{-4mm}
\end{figure}

Fig. \ref{Fig:results_EE} shows the energy consumption ratio, defined as $\eta = {E_{\mathrm{total}}^{\gls{proposed2}}(V_{\mathrm{th}}^{\mathrm{opt}})}/{E_{\mathrm{Baseline}}}$, against $N$. Note that the energy consumption ratio becomes lower than 1 if the energy consumption of \gls{proposed2} is smaller than that of the baseline scheme. From Fig.~\ref{Fig:results_EE}, first, we can see that, when the value of $N$ is smaller (e.g., $N < 5$), $\eta$ is higher than 1, i.e., the total energy consumption of \gls{proposed2} scheme is larger than that of the baseline. This is because the proposed scheme consumes a large amount of energy for receiving the Tiny\gls{ml} model and extracting the similarity measure to suppress a smaller number of image transmissions, while the baseline scheme only transmits a small amount of stored images without consuming energy for Tiny\gls{ml} model introduction. However, as the value of $N$ becomes larger, we can see that the proposed scheme outperforms the baseline in terms of total energy consumption while satisfying the constraint of retrieval accuracy. This is because the \gls{proposed2} can suppress wasteful image transmissions, which are unrelated to the query image, thanks to the introduction of the task-related Tiny\gls{ml} model, leading to a reduction of the overall energy consumption, while the baseline scheme needs to consume substantial energy for the image transmissions. Next, we can see that $\eta$ becomes lower as the value of $\delta$ increases. This is because when the value of $\delta$ becomes higher, the \gls{es} can set a relatively larger $V_{\mathrm{th}}$, by which the \gls{iot} device can suppress the transmission of a large portion of images that are not relevant for the query image. Finally, compared with the results for $b_{q} =8$ and $b_{q} =16$, we can see that the gain of \gls{proposed2} for $b_{q} =8$ is higher (lower) than for $b_{q} =16$ for the range of smaller (higher) value of $N$. For the smaller range of $N$, most of the energy is consumed for receiving the Tiny\gls{ml} model and its processing, for which smaller quantization like $b_{q} =8$ is preferable to reduce these costs. On the other hand, for the larger range of $N$, reducing the transmission cost becomes important for the overall energy reduction, in which case $b_q = 16$ is preferable as it enables the \gls{es} to set a relatively large $V_{\mathrm{th}}$.

\section{Conclusions}
In this letter, focusing on wireless image retrieval from the single \gls{iot} device, we have investigated how we can reduce overall energy consumption when a specific query image is given. In order to reduce wasteful data transmission from the \gls{iot} devices that do not observe the desired image, we have proposed \gls{proposed2}, in which a Tiny\gls{ml} model is transmitted from the \gls{es} to the \gls{iot} device, filtering out undesired image transmission. We have derived a theoretical equation expressing total energy consumption and retrieval accuracy of \gls{proposed2} in order to analyze the performance and effectiveness of our proposed scheme. Our numerical results have revealed that the proposed \gls{proposed2} achieves high energy efficiency while maintaining high retrieval accuracy, especially when the number of stored images is large. 

This work demonstrated the effectiveness of our scheme based on a theoretical concept, which we deem to be sufficient at this initial stage and, based on the promising results, we will carry out experimental evaluation in our future work.  More generally, beyond image retrieval, this analysis can be generalized to encompass a broader set of scenarios in which a preemptive transmission of an ML model can make the subsequent communication more efficient.
\bibliographystyle{IEEEtran}
\bibliography{IEEEabrv,Ref_2}


\end{document}